\documentclass[10pt,twocolumn,superscriptaddress,pra]{revtex4}
\usepackage[latin9]{inputenc}
\usepackage{amsmath}
\usepackage{amssymb}
\usepackage{graphicx}

\makeatletter

\usepackage{amsfonts}
\usepackage{graphicx}
\usepackage{graphics}
\usepackage[usenames]{color}

\makeatother

\begin{document}
\title{Pseudopotentials for Two-dimentional Ultracold Scattering in the Presence
of Synthetic Spin-orbit-coupling}
\author{Christiaan R. Hougaard, Brendan C. Mulkerin, Xia-Ji Liu, Hui Hu, Jia
Wang}
\affiliation{Centre for Quantum and Optical Science, Swinburne University of Technology,
Melbourne 3122, Australia.}
\begin{abstract}
We derive a pseudopotential in two dimensions (2D) with the presence
of a 2D Rashba spin-orbit-coupling (SOC), following the same spirit
of frame transformation in {[}Phys. Rev. A 95, 020702(R) (2017){]}.
The frame transformation correctly describes the non-trivial phase
accumulation and partial wave couplings due to the presence of SOC
and gives rise to a different pseudopotential than the free-space
one, even when the length scale of SOC is significantly larger than
the two-body potential range. As an application, we apply our pseudopotential
with the Lippmann-Schwinger equation to obtain an analytical scattering
matrix. To demonstrate the validity, we compare our results with a
numerical scattering calculation of finite-range potential and shows
perfect agreement over a wide range of scattering energy and SOC strength.
Our results also indicate that the differences between our pseudopotential
and the original free-space pseudopotential are essential to reproduce
scattering observables correctly.
\end{abstract}
\maketitle
Modeling the fundamental two-body interactions is one of the critical
steps in investigating the complex quantum physics of a many-body
system. In particular, for systems with short-range interactions at
low energies such as ultracold quantum gases, the two-body interaction
can be replaced by a zero-range pseudopotential, giving the same wavefunction
outside the original potential. One only needs energy-dependent scattering
lengths obtained via partial expansion of two-body scattering to characterize
the strength of such pseudopotential. For example, in many cases,
s-wave scattering dominates at near-zero temperature, and the Fermi
pseudopotential \citep{Fermi1934,HuangYangPR1957} gives a highly
accurate description of the behavior of degenerated quantum gases.
For higher temperature beyond the Wigner-threshold regime, generalizing
the Fermi pseudopotential with an energy-dependent s-wave scattering
length can give quantitative descriptions \citep{JuliennePRA2002},
when the contribution from the higher-partial wave is negligible.
However, further generalization is necessary near resonances of higher-partial
waves, vanishes of s-wave scattering or when spin-orbit effects couples
higher partial waves. These regimes become particularly interesting
in the field of quantum gases, where interactions can be engineered
at will via Feshbach resonances \citep{ChinRMP2010} and synthetic
spin-orbit-couplings (SOC) \citep{SpielmanNature2011}. Even back
at the initial development of pseudopotential in the 1950s, Yang and
Huang have already made an early attempt to tackle the generalization
for higher partial waves \citep{HuangYangPR1957}. However, they made
an algebraic mistake in their original work leading to an incorrect
prefactor that is discovered and corrected much later \citep{IvanPRL2005,DereviankoPRA2005,TommasoPRL2006}.
With the corrected prefactor, the mean-field energy shift of interacting
fermions in a trap accurately matches experimental measurements \citep{RothPRA2001}.

These pioneer works mentioned so far all focused on pseudopotentials
in three-dimensional (3D) space. Extensions to lower dimensions have
been only of academic interest until an exciting development in the
area of ultracold quantum gases recently: the creation of low-dimensional
quantum gases. Confining quantum gases into lower dimensions can strongly
enhance the quantum correlation of the system, leading to qualitative
changes \citep{BlochRMP2008}. Experimental realization of quasi-one-dimensional
(quasi-1D) bosons allowed the verification of the fermionization of
1D Bose gases in the Tonks--Girardeau regime \citep{SpielmanNature2004,WeissScience2004}.
More recently, tightly confining the motion of atoms in one direction
creates quasi-two-dimensional (quasi-2D) quantum gases that have several
applications in investigating intrigued physics, such as the observation
of the BKT phase \citep{DalibardNature2006,PhillipsPRL2009}, measurement
of the equation of state \citep{RathPRA2010} and super-Efimov physics
\citep{NishidaPRL2013,GaoPRA2015}. The experimental realization of
quasi-2D quantum gas motivates an elegant derivation of 2D pseudopotential
for arbitrary partial waves \citep{BlumePRA2006}.

Another invaluable development in ultracold quantum gases in recent
years is the realization of synthetic gauge fields that can be used
to simulate electromagnetic interactions in systems of neutral particles.
Artificial gauge fields can also couple a particle\textquoteright s
canonical momentum with its (pseudo)spin degrees of freedom \citep{SpielmanNature2011,Zhangjing2014ReviewSOC,Zhaihui2015ReviewSOC},
providing an essential ingredient, namely SOC, for the study of topological
insulators \citep{DalibardRMP2011,Goldman2014RPP}. The interplay
between SOC and short-range interactions might lead to new quantum
behaviors and phases, and soon attracts a lot of interest. In cold-atom
systems, several experimental techniques have been developed to realize
SOC such as lattice shaking \citep{Struck2012PRL} and Raman coupling
\citep{SpielmanNature2011}. While the Raman laser scheme has already
achieved one-dimensional SOC (an equal mixture of Rashba and Dresselhaus
SOC) \citep{SpielmanNature2011,Cheuk2012PRL,Wang2012PRL,Zhang2012PRL,Chunlei2013PRA,Khamehchi2017PRL},
SOC with higher symmetry such as 2D and 3D isotropic ones are more
closely related to the cases in condensed-matter physics. On the other
hand, 2D Rashba (isotropic) SOC, which is our main focus in this work,
is more experimentally accessible than 3D with less problem from heating
\citep{Huang2016NatPhys,Meng2016PRL}.

In previous theoretical studies, people usually directly apply the
original free-space pseudopotentials (obtained from two-body scattering
without SOC) in the presence of SOC \citep{Zhangjing2014ReviewSOC,Zhaihui2015ReviewSOC}.
The justification base on the argument that the characteristic wave-length
of synthetic SOC, in reality, is much larger than the inverse of the
range of short-range interaction. Therefore the effects of SOC were
assumed to have no impact on behaviors of wavefunctions at short inter-particle
distances, and hence the original pseudopotential remains valid from
a perturbation point of view. Nevertheless, in a foresighted study,
Cui points out that the presence of SOC at short distances intrinsically
mixes different partial waves via the couplings of spin, which might
lead to non-trivial influence on the short-range wavefunction \citep{Cui2012PRA}.
Via some numerical investigations, Cui concludes that free-space pseudopotentials,
especially for higher partial waves such as $p$-wave, are not satisfactory.
Ever since, several studies have carefully calculated two-body scattering
with the presence of 3D \citep{ZhangpengPRA2012,YuzhenhuaPRA2012,ZhangpengPRA2013,DuanhaoPRA2013,WangsujuPRA2015,GuanQPRA2016,WangjiaPRA2018}
or 2D \citep{ZhangweiPRA2012,ZhanglongPRA2013} SOC, paving the way
for designing a pseudopotential model. One particularly enlightening
study carried out by Guan and Blume reveals that a frame transformation
approach (that we will detail later) can correctly calculate the scattering
phase accumulated at short distances modified by SOC \citep{GuanQPRA2017}.
However, a proper pseudopotential that includes the nonperturbative
effects of SOC at short-range and correctly reproduces scattering
observables is still missing. In this Rapid Communication, we derive
an analytical form of the pseudopotential in 2D with the presence
of 2D Rashba SOC, following the same spirit of frame transformation
in Ref. \citep{GuanQPRA2017}. To verify the validity, we apply the
Lippmann-Schwinger equation to obtain the analytical scattering matrix
and compare it with a numerical scattering calculation with finite-range
potential.

We first give a brief review of 2D pseudopotential in the free-space
without the presence of SOC. We consider two identical particles ($n=1,2$)
of mass $m$ confined in a 2D $x$-$y$ plane with position vectors
$\mathbf{r}_{n}$. Seperating out the center-of-mass (COM) motion,
the Hamiltonian of the relative motion is given by $H^{{\rm fs}}=\mathbf{p}^{2}/2\mu_{2b}+U(\rho)$,
where $\mu_{2b}=m/2$ is the two-body reduced mass, $\mathbf{r}=\{\rho,\phi\}$
is the relative position in polar coordinates, and $\mathbf{p}=-i\hbar\{\partial_{\rho},\rho^{-1}\partial_{\phi}\}$
is the relative momentum in 2D. We also assume the potential $U(\rho)$
is isotropic and short-range, i.e., vanishes beyond a small radius
$\rho_{0}$. The isotropic symmetry allows the wavefunction to be
expanded as $\Psi^{{\rm fs}}(\mathbf{r})=\sum_{m_{\ell}}R_{m_{\ell}}^{{\rm fs}}(\rho)\Phi_{m_{\ell}}(\phi)$,
where $\Phi_{m_{\ell}}(\phi)=e^{im_{\ell}\phi}/\sqrt{2\pi}$, $R_{m_{\ell}}^{{\rm fs}}(\rho)$
satisfies the radial Schr{\"o}dinger equation
\begin{equation}
\left[-\frac{\hbar^{2}}{2\mu_{2b}}\left(\frac{\partial^{2}}{\partial\rho^{2}}+\frac{1}{\rho}\frac{\partial}{\partial\rho}-\frac{m_{\ell}^{2}}{\rho^{2}}\right)+U(\rho)-E\right]R_{m_{\ell}}^{{\rm fs}}(\rho)=0\label{eq:fsEqSchrodinger}
\end{equation}
and adopts an asymptotic form $R_{m_{\ell}}^{{\rm fs}}(\rho)\propto J_{m_{\ell}}(k\rho)-\tan[\delta_{m_{\ell}}(k)]N_{m_{\ell}}(k\rho)$
for $\rho>\rho_{0}$. Here $E=\hbar^{2}k^{2}/2\mu_{2b}$ and $J_{m_{\ell}}(k_{\tau}r)$
and $Y_{m_{\ell}}(k_{\tau}r)$ are the Bessel functions of the first
and second kind respectively. $\delta_{m_{\ell}}(k)$ are the energy-dependent
phase shifts, satisying threshold law $\tan[\delta_{0}(k)]\propto1/\log k$
and $\tan[\delta_{m_{\ell}}(k)]\propto1/k^{2|m_{\ell}|}$ for $|m_{\ell}|\ge1$.
Reference \citep{BlumePRA2006} shows that replacing $U(\rho)$ by
a pseudopotential $V_{m_{\ell}}^{{\rm fs}}(\rho)$ can give the same
asymptotic wavefunction, and hence reproduce the low-energy observables
of the original finite-range potential. The explicit form of $V_{m_{\ell}}^{{\rm fs}}(\rho)$
in free space is given by
\begin{equation}
V_{m_{\ell}}^{{\rm fs}}(\rho,k)=-\frac{\hbar^{2}}{\mu_{2b}}\frac{\tan\left[\delta_{m_{\ell}}(k)\right]}{c_{m_{\ell}}k^{2m_{\ell}}\rho^{m_{\ell}}}\left[\frac{\delta(\rho-s)}{2\pi\rho}\hat{O}_{m_{\ell}}(\rho,k)\right]_{s\rightarrow0^{+}},\label{eq:fspseudopotential}
\end{equation}
where $c_{m_{\ell}}=(2m_{\ell})!/[\Gamma(m_{\ell}+1)]^{2}2^{2m_{\ell}}$
with $\Gamma(\cdot)$ being the gamma function. The form of delta
shell $\delta(\rho-s)$ of radius $s$ approaches to a contact potential
$\delta(\rho)$ in the limit $s\rightarrow0$, and allows us to deal
with the divergence of the regularized operator rigorously. The regularized
operator reads as

\begin{equation}
\hat{O}_{m_{\ell}}=\left\{ \begin{array}{ll}
{\frac{2}{1-\tan\left[\delta_{0}(k)\right]f_{0}(k,\rho)}\frac{\partial}{\partial\rho}\rho} & {;m_{\ell}=0}\\
{\frac{2}{1-\tan\left[\delta_{m_{\ell}}(k)\right]f_{m_{\ell}}(k,\rho)}\frac{\partial^{2m_{\ell}}}{\partial\rho^{2m_{\ell}}}\rho^{m_{\ell}}} & {;m_{\ell}>0}
\end{array}\right.\label{eq:regularizedoperator}
\end{equation}
where 
\begin{equation}
f_{m_{\ell}}(k,\rho)=\left\{ \begin{array}{ll}
{\frac{2}{\pi}\left[1+\gamma+\log\left(\frac{1}{2}k\rho\right)\right]} & {;m_{\ell}=0}\\
{\frac{2}{\pi}\left[\sum_{n=0}^{2m_{\ell}-1}\frac{1}{2m_{\ell}-n}-\bar{\psi}+\log\left(\frac{1}{2}k\rho\right)\right]} & {;m_{\ell}>0}
\end{array}\right..
\end{equation}
Here $2\bar{\psi}(m_{\ell})=\psi(1)+\psi(m_{\ell}+1)$, where $\psi$
denotes the digamma function. For $m_{\ell}<0$, the pseudopotential
in Eq. (\ref{eq:fspseudopotential}) takes the same form but with
$m_{\ell}$ replaced by $|m_{\ell}|$. In contrast to the 3D pseudopotential,
the $\tan\left[\delta_{m_{\ell}}(k)\right]$ dependence in the denominator
of the regularized operators originates from the fact that $\frac{\partial}{\partial\rho}\rho N_{0}(k\rho)$
and $\frac{\partial^{2m_{\ell}}}{\partial\rho^{2m_{\ell}}}\rho^{m_{\ell}}N_{m_{\ell}}(k\rho)$
does not vanishes at $\rho\rightarrow0$.

Now we consier the effects of SOC, where each particle feels a 2D
Rashba SOC described by $H_{{\rm SO}}^{(n)}=k_{{\rm SO}}\mathbf{p}_{n}\cdot\mathbf{s}_{n}/m$,
with $\mathbf{p}_{n}$ and $\mathbf{s}_{n}$ being the 2D momentum
and spin operator of particle $n$ respectively. Following the spirit
of Refs. \citep{DuanhaoPRA2013,WangsujuPRA2015,GuanQPRA2016,GuanQPRA2017,WangjiaPRA2018},
we focus on the scattering in the COM frame, where the relative Hamiltonian
can be written as $H_{{\rm rel}}=H^{{\rm fs}}+V^{{\rm SO}}$ with
$V^{{\rm SO}}=k_{{\rm SO}}\mathbf{\Sigma}\cdot\mathbf{p}/2\mu_{2b}$
describing the SOC effect. and $\mathbf{\Sigma}=\mathbf{s_{1}-\mathbf{s_{2}}}$
is the relative spin operator. $k_{{\rm SO}}$ defines the strength
of SOC coupling, and gives an energy scale $E_{{\rm SO}}=\hbar^{2}k_{{\rm SO}}^{2}/2m$.

A formal way to solve the corresponding relative Schr{\"o}dinger equation
is to formulate it as a multichannel problem by expanding the $\tau$'th
independent solution as
\begin{equation}
\Psi_{\tau}^{{\rm SO}}(\mathbf{r})=\sum_{\nu}R_{\nu\tau}^{{\rm SO}}(\rho)A_{\nu}(\Omega),
\end{equation}
where the channel functions $A_{\nu}(\Omega)\equiv\langle\Omega|\nu\rangle$
are functions of $\Omega$ that includes all degrees of freedom except
for $\rho$. Due to the azimutual symmetry, total angular momentum
(along $z$-axis) $m_{j}$ is a good quantum number that equals to
$m_{\ell}+m_{S}$. Here $m_{1}$, $m_{2}$ and $m_{S}=m_{1}+m_{2}$
are quantum number of the projection of the operator $\mathbf{s}_{1}$,
$\mathbf{s}_{2}$ and $\mathbf{S}=\mathbf{s}_{1}+\mathbf{\mathbf{s}}_{2}$
to the quantization $z$-axis respectively. Defining the total spin
basis $\left|\chi\right\rangle \equiv\left|(s_{1,}s_{2}),S,m_{S}\right\rangle $
as usual, we choose the channel functions being $A_{\nu}(\Omega)=i^{m_{\ell}}\Phi_{m_{\ell}}(\phi)\left|\chi\right\rangle $
with $m_{\ell}+m_{S}=m_{j}$ and $S+m_{\ell}+s_{1}+s_{2}$ being even/odd
for bosons/fermions respectively. The subindex $\nu$ collectively
represents quantum numbers $\{m_{\ell},S,m_{S};m_{j}\}$. Here we
omit quantum numbers $s_{1}$ and $s_{2}$ in the channel index notations
since they are the same for all channels. At $\rho>\rho_{{\rm 0}},$
wavefuctions can be expressed as a linear combination of non-interacting
(but with SOC) regular and irregular solution $\underline{R}^{{\rm SO}}=\underline{F}-\underline{G}\underline{\mathcal{K}}$,
where $\underline{R}^{{\rm SO}}$ is the matrix form of raidal solutions
$R_{\nu\tau}^{{\rm SO}}(\rho)$. (Through out this paper, underline
implies matrix form.) The matrix elements of regular solution $\underline{F}$
can be written as $F_{\nu\tau}(\rho)=N_{\tau}C_{\nu\tau}\sqrt{k_{\tau}}J_{m_{\ell}}(k_{\tau}\rho)$,
where $C_{\nu\tau}$, $k_{\tau}$ and $N_{\tau}$ can be obtained
by diagonalizing the non-interacting Hamiltonian using the same procedure
as Ref. \citep{WangsujuPRA2015}. The corresponding irregular solution
can be obtained as $G_{\nu\tau}(\rho)=N_{\tau}C_{\nu\tau}\sqrt{k_{\tau}}Y_{m_{\ell}}(k_{\tau}\rho)$.
The scattering $K$ matrix $\underline{\mathcal{K}}$ determines scattering
observables and is related to the more familiar $S$ matrix by $\underline{\mathcal{S}}=(\underline{I}+i\underline{\mathcal{K}})(\underline{I}-i\underline{\mathcal{K}})^{-1}$.
Our goal is to replace the potential $U(\rho)$ by a potential $\underline{V}^{(m_{j})}(\rho)$
that acts only at $\rho=0$, and gives the same asymptotic wavefunction
and hence the same $K$ matrix. Here the underline indicates $\underline{V}^{(m_{j})}(\rho)$
is a matrix and not nessesary diagonal due to the presene of SOC.

To derive this pseudopotential, we follow the spirit of Ref. \citep{GuanQPRA2017},
and apply a frame transformation approach. Defining a unitary transformation
$\mathcal{U}_{1}=\exp(-ik_{{\rm SO}}\mathbf{\Sigma}\cdot\mathbf{r}/2\hbar)$,
the ``rotated'' Hamiltonian $H^{{\rm temp}}\equiv\mathcal{U}_{1}^{-1}H_{{\rm rel}}\mathcal{U}_{1}=H^{{\rm fs}}+\mathcal{\epsilon^{{\rm temp}}}+\mathcal{O}(\rho)$
is introduced as an intermediate step. Here we neglect terms of order
$\rho$ and higher denoted by $\mathcal{O}(\rho)$, since the pseudopotential
will only act at $\rho=0$. The constant term $\mathcal{\epsilon^{{\rm temp}}}$
is given by $\mathcal{\epsilon^{{\rm temp}}}=-E_{{\rm SO}}(\Sigma_{x}^{2}+\Sigma_{y}^{2}+S_{z}L_{z})/2\hbar^{2}$,
where $\Sigma_{x}$ ($\Sigma_{y}$) are the $x$($y$)-component of
$\mathbf{\Sigma}$, $S_{z}$ is the $z$-component of total spin operator
$\mathbf{S}$, and $L_{z}=-i\hbar\partial_{\phi}$ is the 2D angular
momentum operator. For two spin-1/2 particles, this operator expanded
by channel functions $A_{\nu}(\Omega)$ gives a diagonal matrix $\mathcal{\underline{\epsilon}^{{\rm temp}}}$
. In contrary, for higher spins, $\mathcal{\underline{\epsilon}^{{\rm temp}}}$
is in general not diagonal, where the only non-zero matrix elements
are the ones couples channels with the same $m_{S}$ and hence the
same $m_{\ell}$. Therefore, one can introduc another $\rho$-indepdent
unitary transformation $\mathcal{U}_{2}$ that is block-diagonal in
$m_{\ell}$ subspaces and satisfies $\underline{\epsilon}=\mathcal{U}_{2}^{-1}\underline{\epsilon}^{{\rm temp}}\mathcal{U}_{2}$
is diagonal. Automatically, $\mathcal{U}_{2}^{-1}H^{{\rm fs}}\mathcal{U}_{2}=H^{{\rm fs}}$
is also diagonal. Therefore, we find a unitary transformation $\mathcal{U}=\mathcal{U}_{2}\mathcal{U}_{1}$
that leads to a set of uncoupled radial Schr{\"o}dinger equation that
is at least valid near the origin $\rho=0$, which is given by,
\begin{equation}
\left[-\frac{\hbar^{2}}{2\mu_{2b}}\left(\frac{\partial^{2}}{\partial\rho^{2}}+\frac{1}{\rho}\frac{\partial}{\partial\rho}-\frac{m_{\ell}^{2}}{\rho^{2}}\right)+U(\rho)-E_{\nu}\right]\tilde{R}_{\nu\nu}^{{\rm SO}}(\rho)=0,
\end{equation}
where $E_{\nu}\equiv E+\epsilon_{\nu}$ with $\epsilon_{\nu}$ being
the diagonal matrix elements of SOC-induced energy shift $\underline{\epsilon}$.
Comparing with the free-space Schr{\"o}dinger equation in Eq. (\ref{eq:fsEqSchrodinger}),
a pseudopotential $\underline{\tilde{V}}^{(m_{j})}(\rho,k)$ with
diagonal matrix elements $\tilde{V}_{\nu\nu}^{(m_{j})}(\rho,k)=V_{m_{\ell}}^{{\rm fs}}(\rho,k_{\nu})$
where $k_{\nu}^{2}/2\mu_{2b}=E_{\nu}$ can reproduce the wavefunction
$\tilde{R}_{\nu\nu}^{{\rm SO}}(\rho)$ in the rotated frame. The pseudopotential
in the original frame can therefore be obtained by an inverse rotation:
\begin{equation}
\underline{V}^{(m_{j})}(\rho,k)=\mathcal{U}\tilde{\underline{V}}^{(m_{j})}(\rho,k)\mathcal{U}^{-1}.\label{eq:pseudopotential}
\end{equation}
Once we obtained $\underline{V}^{(m_{j})}(\rho,k)$ for all $m_{j}$
(or up to a cut-off $m_{j}$ in practice), the total pseudopotential
can be expressed as
\begin{equation}
\begin{gathered}V(\rho,k)\Psi^{{\rm SO}}(\mathbf{r})=\sum_{m_{j}}\sum_{\nu\nu'}A_{\nu}(\Omega)\int\rho'd\rho'\underline{V}_{\nu\nu'}^{(m_{j})}(\rho')\\
\times\int d\Omega'A_{\nu'}^{*}(\Omega')\Psi^{{\rm SO}}(\mathbf{r}')
\end{gathered}
,
\end{equation}
which might be helpful for applications where $m_{j}$ is not a good
quantum number.

For illustration, we consider two spin-$1/2$ fermions in the $m_{j}=0$
subspace, and omit the notation of $m_{j}$ hereafter unless specify
otherwise. The basis are denoted by $\nu\equiv\{m_{\ell},S,m_{S}\}=\{-1,1,1\},\{0,0,0\},\{1,1,-1\}$.
In this order of the basis, the rotational matrix can be written out
explicitely
\begin{equation}
\underline{\mathcal{U}}=\left[\begin{array}{ccc}
{\cos^{2}\left(\frac{\lambda\rho}{2}\right)} & {-\frac{\sin(\lambda\rho)}{\sqrt{2}}} & {-\sin^{2}\left(\frac{\lambda\rho}{2}\right)}\\
{\frac{\sin(\lambda\rho)}{\sqrt{2}}} & {\cos(\lambda\rho)} & {\frac{\sin(\lambda\rho}{\sqrt{2}}}\\
{-\sin^{2}\left(\frac{\lambda\rho}{2}\right)} & {-\frac{\sin(\lambda\rho)}{\sqrt{2}}} & {\cos^{2}\left(\frac{\lambda\rho}{2}\right)}
\end{array}\right],
\end{equation}
where $\lambda=k_{{\rm SO}}/2$ is introduced for convenience. After
rotation, the SOC-induced energy shift is given by $\underline{\epsilon}={\rm diag}[0,\hbar^{2}\lambda^{2}/\mu_{2b},0]$,
where ${\rm diag[\cdot]}$ represents a diagonal matrix. The energy
shift determines the pseudopotential in the rotated frame as $\tilde{\underline{V}}(\rho,k)={\rm diag}[V_{1}^{{\rm fs}}(\rho,k_{p}),V_{0}^{{\rm fs}}(\rho,k_{s}),V_{1}^{{\rm fs}}(\rho,k_{p})]$,
where $k_{s}=\sqrt{k^{2}+2\lambda^{2}}$ and $k_{p}=k$. We then apply
Eq. (\ref{eq:pseudopotential}) to obtain the pseudopotential in the
original frame. We can write the the pseudopotential as a summation
of $s$- and $p$- wave contribution:

\begin{equation}
\underline{V}=-\frac{\hbar^{2}}{\mu_{2b}}\left\{ \frac{\delta(\rho-s)}{2\pi\rho}\left[\tan\delta_{s}(k_{s})\underline{O}_{s}+\frac{\tan\delta_{p}(k_{p})}{k_{p}^{2}}\underline{O}_{p}\right]\right\} _{s\rightarrow0},
\end{equation}
where

\begin{equation}
\underline{O}_{s}=\left[\begin{array}{ccc}
0 & 0 & 0\\
-\frac{\lambda}{\sqrt{2}}\hat{O}_{0}\rho & \frac{1}{\rho}\hat{O}_{0} & -\frac{\lambda}{\sqrt{2}}\hat{O}_{0}\rho\\
0 & 0 & 0
\end{array}\right],
\end{equation}
and

\begin{equation}
\underline{O}_{p}=\left[\begin{array}{ccc}
\frac{1}{\rho}\hat{O}_{1}\left(1-\frac{\lambda^{2}\rho^{2}}{4}\right) & \frac{\lambda}{\rho\sqrt{2}}\hat{O}_{1}\rho & -\frac{1}{\rho}\hat{O}_{1}\frac{\lambda^{2}\rho^{2}}{4}\\
\frac{\lambda}{\sqrt{2}}\hat{O}_{1}\left(1-\frac{\lambda^{2}\rho^{2}}{2}\right) & \lambda^{2}\hat{O}_{1}\rho & \frac{\lambda}{\sqrt{2}}\hat{O}_{1}\left(1-\frac{\lambda^{2}\rho^{2}}{2}\right)\\
-\frac{1}{\rho}\hat{O}_{1}\frac{\lambda^{2}\rho^{2}}{4} & \frac{\lambda}{\rho\sqrt{2}}\hat{O}_{1}\rho & \frac{1}{\rho}\hat{O}_{1}\left(1-\frac{\lambda^{2}\rho^{2}}{4}\right)
\end{array}\right].
\end{equation}

\begin{figure}
\includegraphics[width=0.98\columnwidth]{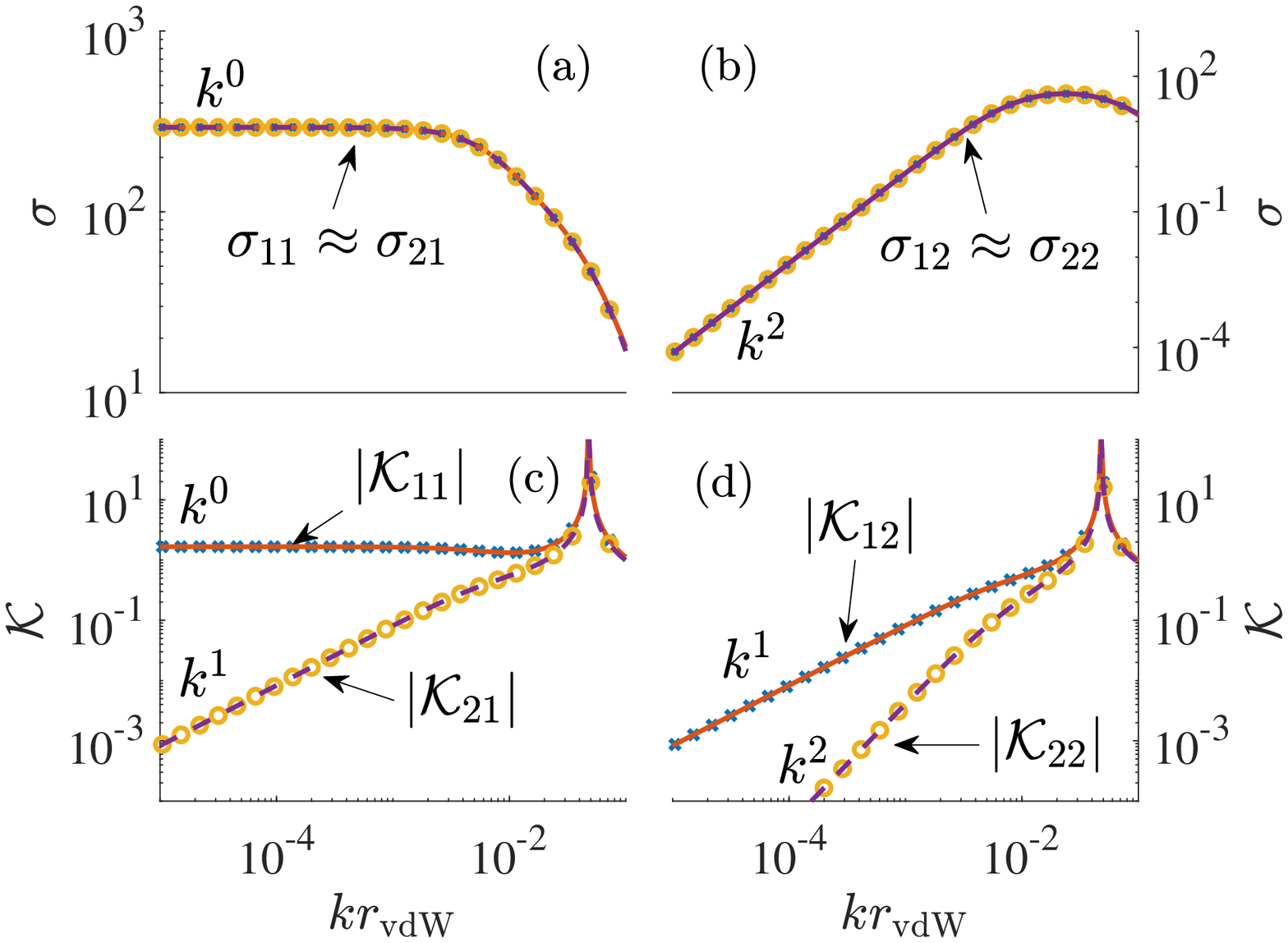}

\caption{Results near $s$-wave resonances with $\lambda r_{{\rm vdW}}=0.01$.
The free-space scattering phase-shifts are obtained from a Lennard-Jones
model potential with parameter $r_{0}=0.58r_{{\rm vdW}}$ that leads
to $a_{s}(0)\approx21.777r_{{\rm vdW}}$. (a) and (b) partial cross-sections.
(c) and (d) $K$ matrix elements. The curves represents analytical
results, and the symbols are obtained by numerical scattering calculations.\label{fig:swavek}}
\end{figure}

\begin{figure}
\includegraphics[width=0.98\columnwidth]{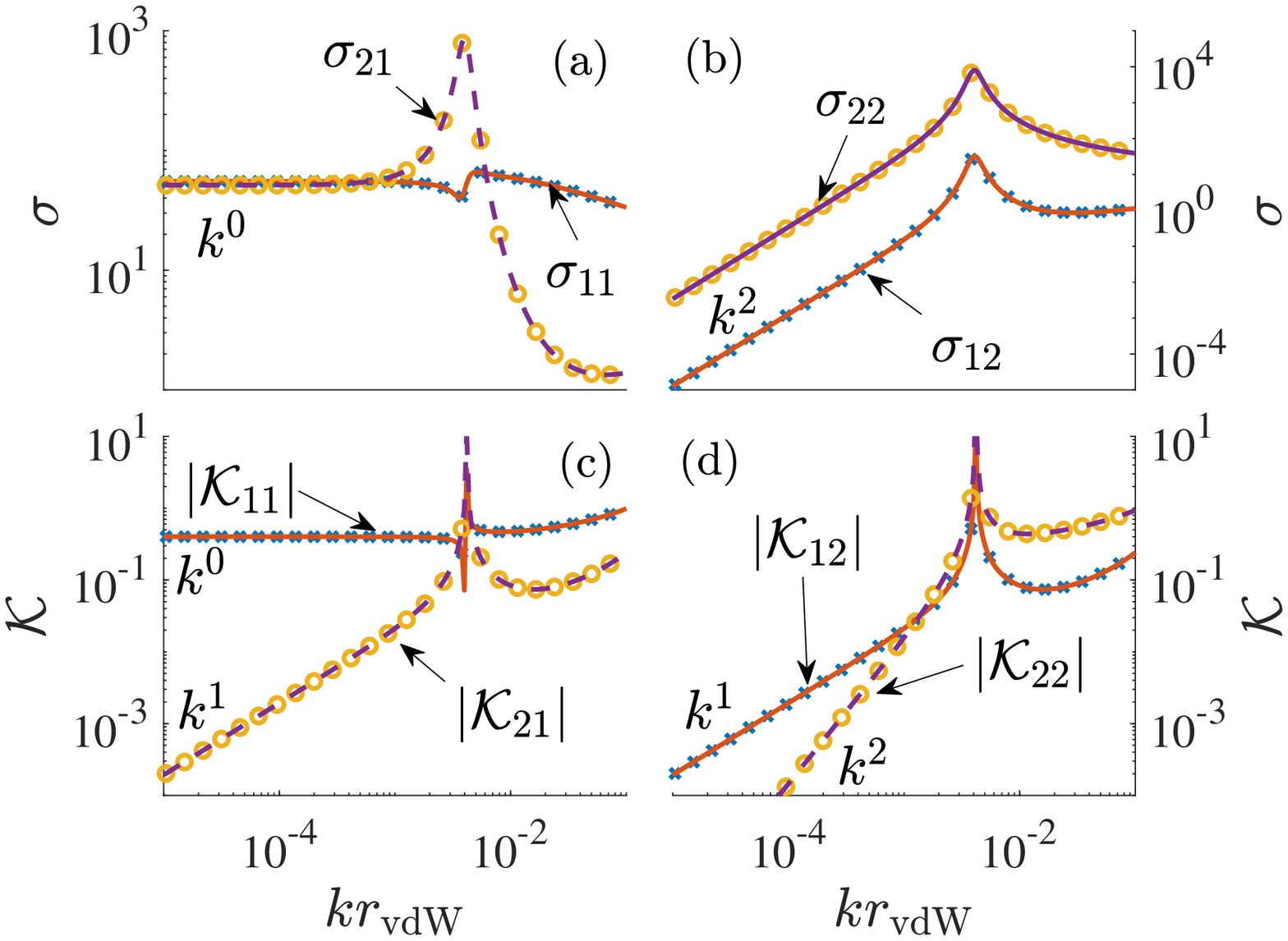}\caption{Results near $p$-wave resonances with $\lambda r_{{\rm vdW}}=0.01$.
The free-space scattering phase-shifts are obtained from a Lennard-Jones
model potential with parameter $r_{0}=0.552981r_{{\rm vdW}}$ that
leads to $A_{p}\approx-8.577\times10^{5}r_{{\rm vdW}}^{2}$. (a) and
(b) partial cross-sections. (c) and (d) $K$ matrix elements. The
curves represents analytical results, and the symbols are obtained
by numerical scattering calculations. \label{fig:pwavek}}
\end{figure}

Terms of higher order of $\rho$ can be ignored with the consideration
that the pseudopotential only contribute to $K$ matrix with terms
proportional to $F_{\nu'\tau'}^{*}(s)V_{\nu\nu'}(s,k)F_{\nu\tau}(s)_{s\rightarrow0}$
and $F_{\nu'\tau'}^{*}(s)V_{\nu\nu'}(s,k)G_{\nu\tau}(s)_{s\rightarrow0}$
{[}see Eqs. (\ref{eq:ABmatrix}) below{]}. Comparing with the the
free-space pseudopotential ${\rm diag}[V_{1}^{{\rm fs}}(k,\rho),V_{0}^{{\rm fs}}(k,\rho),V_{1}^{{\rm fs}}(k,\rho)]$,
there are two important differences. One is the SOC-induced energy
shift leads to a different $s$-wave phase shift $\delta_{s}(k_{s})$.
The other is the non-diagonal terms rised from the rotational transformation,
which describes the intrinsical partial waves mixing at short distances
induced by SOC. As we will see, both of these differences play significant
roles in producing scattering observables correctly.

To verify the validity of the pseudopotential, we apply the Lippmann-Schwinger
equation to calulate the $K$ matrix. The Lippmann-Schwinger equation
is the integral form of the Schr{\"o}dinger equation $\Psi_{\tau}(\mathbf{r})=\Psi_{0}(\mathbf{r})+\int G(\mathbf{r},\mathbf{r'})V(\mathbf{r'})\Psi_{\tau}(\mathbf{r}')d\mathbf{r}'$,
or equivalently in the matrix form
\begin{equation}
\underline{R}^{{\rm SO}}(\rho)=\underline{F}(\rho)+\int\underline{\mathcal{G}}(\rho,\rho')\underline{V}(\rho')\underline{R}(\rho')\rho'd\rho'.
\end{equation}
Here $\underline{\mathcal{G}}(\rho,\rho')$ is the matrix representation
of the Green's function $\mathcal{G}(\mathbf{r},\mathbf{r'})=\sum_{\nu\nu'}A_{\nu}(\Omega)\underline{\mathcal{G}}_{\nu\nu'}(\rho,\rho')A_{\nu'}^{*}(\Omega)$,
which is given by
\begin{equation}
\underline{\mathcal{G}}\left(\rho,\rho^{\prime}\right)=\pi\left\{ \begin{array}{ll}
{\underline{F}(\rho)G^{\dagger}\left(\rho^{\prime}\right),} & {\rho<\rho^{\prime}}\\
{\underline{G}(\rho)\underline{F}^{\dagger}\left(\rho^{\prime}\right),} & {\rho>\rho^{\prime}}
\end{array}.\right.
\end{equation}
The Green's function approach becomes very helpful when the potential
can be replaced by a pseudopotential $\underline{V}(\rho')\propto\delta(\rho')$,
and the $K$ matrix can then be obtained by $\underline{\mathcal{K}}=(I+\underline{\mathcal{B}})^{-1}\mathcal{\underline{A}}$,
where
\begin{equation}
\begin{gathered}\underline{\mathcal{A}}=-\pi\int\rho'd\rho'\underline{F}(\rho')\underline{V}(\rho')\underline{F}(\rho'),\\
\underline{\mathcal{B}}=-\pi\int\rho'd\rho'\underline{F}(\rho')\underline{V}(\rho')\underline{G}(\rho').
\end{gathered}
\label{eq:ABmatrix}
\end{equation}
Noticing it is a special property of 2D that $\mathcal{B}$ does not
vanish. As a direct consequence, the $K$ matrix in general cannot
be written as a summation of $s$- and $p$-wave contribution in contrast
to the 3D case as shown in Eq. (11) of Ref. \citep{GuanQPRA2017}.

For illustration, we focus on the case of two spin-$1/2$ fermions
in the $m_{j}=0$ subspace. The regular solution $\underline{F}$
and irregular solution $\underline{G}$ can be determined by the coefficient
$C_{\nu\tau}$ in a matrix form
\begin{equation}
\underline{C}=\left[\begin{array}{ccc}
{-1/2} & {-1/2} & {1/\sqrt{2}}\\
{-1/\sqrt{2}} & {1/\sqrt{2}} & {0}\\
{1/2} & {1/2} & {1/\sqrt{2}}
\end{array}\right],
\end{equation}
where the the column index $\tau$ corresponding to cannonical momentum
$\{k_{\tau}\}\equiv\{k_{1},k_{2},k_{3}\}=\{k_{b}+\lambda,k_{b}-\lambda,k\}$
and normalization $\{\hbar^{2}N_{\tau}^{2}/\mu_{2b}\}=\left\{ 1/k_{b},1/k_{b},1/k\right\} $
where $k_{b}=\sqrt{\lambda^{2}+k^{2}}$. One can identify $\tau=\{1,2,3\}$
corresponds to three different configurations $|-,-\rangle$, $|+,+\rangle$
and $|-,+\rangle$, where $-$ ($+$) indicates the helicity, i.e.
whether the spin is anti-parallel/parallel to the direction of current
\citep{DuanhaoPRA2013}.

Inserting $\underline{F}$ and $\underline{G}$ into Eq. (\ref{eq:ABmatrix})
gives $\underline{\mathcal{A}}$ and $\underline{\mathcal{B}}$ that
determines $\underline{\mathcal{K}}$. We find that the $K$ matrix
is block-diagonal and can be expressed as
\begin{equation}
\mathcal{\underline{K}}=\left[\begin{array}{cc}
\underline{\mathcal{K}}^{(+)} & 0\\
0 & \underline{\mathcal{K}}^{(-)}
\end{array}\right],
\end{equation}
where $\underline{\mathcal{K}}^{(-)}=\tan[\delta_{p}(k_{p})]$, and
$\underline{\mathcal{K}}^{(+)}$ is a $2\times2$ matrix. The block-diagonal
structure can be understood by studying the $\mathcal{\mathcal{PT}}$-symmetry
of $\sum_{\nu}F_{\nu\tau}(\rho)A_{\nu}(\Omega)$ in the $m_{j}=0$
subspace, where $\mathcal{P}$ is defined as $\phi\rightarrow\phi+\pi$
and $\mathcal{T}$ is defined as $\left|s_{n},m_{n}\right\rangle \rightarrow\left|s_{n},-m_{n}\right\rangle $
for both $n=1,2$. Defining $\mathcal{P}\mathcal{T}[\sum F_{\nu\tau}(\rho)A_{\nu}(\Omega)]=\Pi_{\tau}[\sum F_{\nu\tau}(\rho)A_{\nu}(\Omega)]$,
one finds that $\Pi_{\tau}=+1$ /$-1$ for $\tau=\{1,2\}$/$\{3\}$,
leading to the block-diagonal structure. {[}For $m_{j}\ne0$ subspaces,
${\rm sign}(m_{j})$ and $\mathcal{P}\mathcal{T}$ cannot simutaneously
be good quantum numbers{]}. The expressions of matrix elements of
$\underline{\mathcal{K}}^{(+)}$ are analytical but quite cumbersome,
and hence we only give the full expression in the supplemental materials
and illustrate them in Fig. \ref{fig:swavek} and Fig. \ref{fig:pwavek}
as two numerical examples near $s$- and $p$-wave resonances respectively.
In these two examples, the energy-dependent phase-shifts $\delta_{s}(k)$
and $\delta_{p}(k)$ are obtained from a free-space scattering calculation
with Lenard-Jones potential $U(\rho)=-\frac{C_{6}}{\rho^{6}}\left(1-\frac{r_{0}^{6}}{\rho^{6}}\right)$,
where $C_{6}$ defines a length scale $r_{{\rm vdW}}=(2\mu_{2b}C_{6}/\hbar^{2})^{1/4}/2$
and $r_{0}$ controls short-range physics and is used to tune zero-energy
scattering phase shifts. The analytical results shows a good agreement
with a full numerical scatteirng calculation with the representation
of SOC, using a similar procedure as Ref. \citep{WangjiaPRA2018}.
The technical details are shown in the Supplemental Material.

\begin{figure}
\includegraphics[width=0.98\columnwidth]{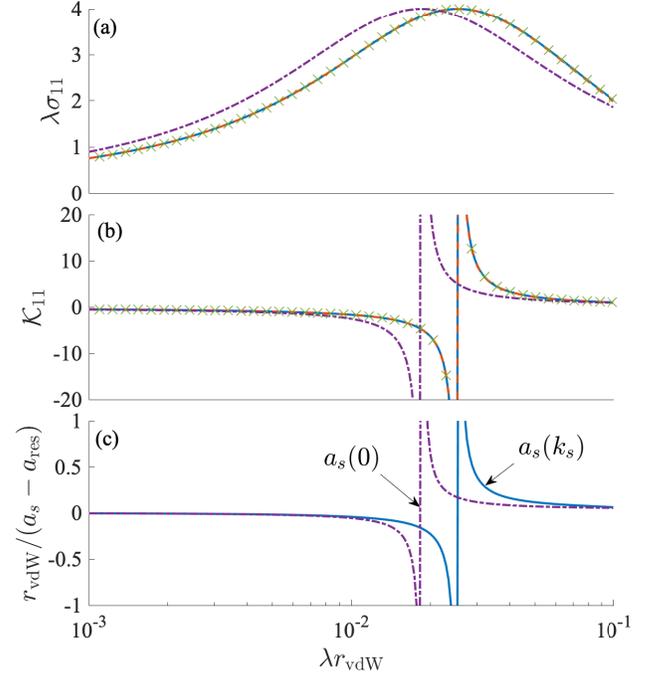}

\caption{Scattering results at zero scattering energy for the same Lennard-Jones
potential of Fig. \ref{fig:swavek}. (a) Scaled partial cross-section
$\lambda\sigma_{11}$ as a function of $\lambda$. (b) $K$ matrix
element $\mathcal{K}_{11}$ as a function of $\lambda$. The blue
solid curves are the analytical results, and the red dashed curves
are determined by the $s$-wave only approximation Eq. (\ref{eq:Ksmat}),
which are indistinguishable to the solid curves on the scale shown.
The purple dash-dotted curves are calculated using the free-space
pseudopotential directly. The green crosses are results from a numerical
calculation using the same Lennard-Jones potential with the presence
of SOC. (c) The blue solid curve shows $r_{{\rm vdW}}/[a_{s}(k_{s})-a_{{\rm res}}]$
as a function of $\lambda$, while the purple dash-dotted curve shows
$r_{{\rm vdW}}/[a_{s}(0)-a_{{\rm res}}]$.\label{fig:swavelambda}}
\end{figure}

Near $s$-wave resonance, $p$-wave scattering is negligible and the
scattering matrix can be simplified as
\begin{equation}
\underline{\mathcal{K}}^{(s)}=T_{0}\left[\begin{array}{cc}
k_{1} & -k\\
-k & k_{2}
\end{array}\right],\label{eq:Ksmat}
\end{equation}
where $T_{0}=\tan[\delta_{s}(k_{s})]/2k_{b}\alpha_{s}$ and $\alpha_{s}=1+\frac{2}{\pi}\tan[\delta_{s}(k_{s})]\left(\log\frac{k}{k_{s}}+\frac{\lambda}{k_{b}}\tanh^{-1}\frac{\lambda}{k_{b}}\right)$.
The $S$ matrix can be obtained via $\underline{\mathcal{S}}^{(s)}=(I+i\mathcal{\underline{\mathcal{K}}}^{(s)})(I-i\mathcal{\underline{K}}^{(s)})^{-1}$
and determines the scattering cross-section $\sigma_{\tau\tau'}^{(s)}=2|\mathcal{\underline{S}}_{\tau\tau'}^{(s)}-\delta_{\tau\tau'}|^{2}/k_{\tau}$
that are read as $\sigma_{11}^{(s)}=\sigma_{21}^{(s)}=k_{1}8T_{0}^{2}/(1+4T_{0}^{2}k_{b}^{2})$
and $\sigma_{12}^{(s)}=\sigma_{22}^{(s)}=k_{2}8T_{0}^{2}/(1+4T_{0}^{2}k_{b}^{2})$.
As a result, $\sigma_{21}^{(s)}/\sigma_{12}^{(s)}=k_{1}/k_{2}>1$
indicates that particles are preferentially scattered into the lower
energy helicity ``$-$'' state. The validity of Eq. (\ref{eq:Ksmat})
near $s$-wave resonances is varified in Fig. \ref{fig:swavek}. In
the zero-energy limit, $\lambda\sigma_{11}^{(s)}\rightarrow4/\left\{ 1+[\gamma+\frac{2}{\pi}\log\left(\lambda a_{s}(\sqrt{2}\lambda)\right)]^{2}\right\} $,
where $a_{s}(k)$ is the generalized energy-dependent $s$-wave scattering
length defined by $\cot[\delta_{s}(k)]=\frac{2}{\pi}\log[k_{s}a_{s}(k)]+\gamma$.
The rescaled cross-section therefore reach maximum when $a_{s}(k_{s})$
equals to $a_{{\rm res}}\equiv e^{-\pi\gamma/2}/\lambda$. In comparison,
if we replace $U(\rho)$ directly by ${\rm diag}[0,V_{0}^{{\rm fs}}(k,\rho),0]$,
the free-space pseudopotential with $s$-wave only, and apply the
Lippmann-Schwinger equation, the obtained $K$ matrix will obey the
same formula Eq. (\ref{eq:Ksmat}) with $k_{s}$ replaced by $k$.
Consequently, the rescaled cross-section reaches maximum when $a_{s}(0)=a_{{\rm res}}$.
Figure \ref{fig:swavelambda} shows such comparison, where one can
see the SOC-induced energy shift that leads to $k_{s}\ne k$ is crucial
to chareterize the two-body scattering correctly, especially near
the maximum of $\lambda\sigma_{11}$.

\begin{figure}

\includegraphics[width=0.98\columnwidth]{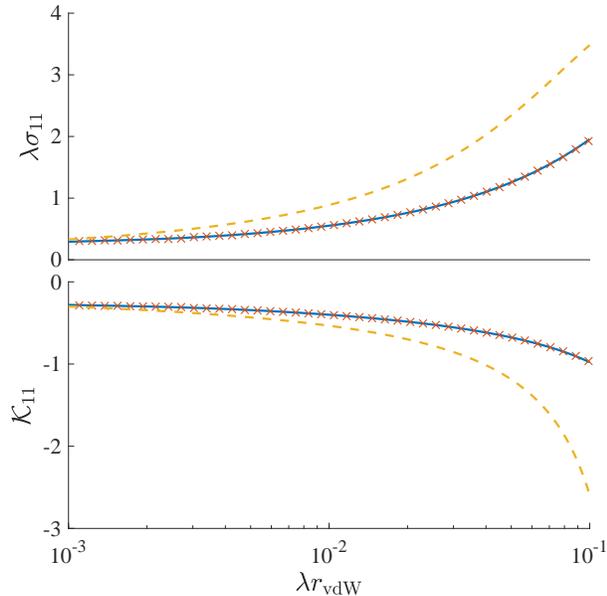}\caption{Scattering results at zero scattering energy for the same Lennard-Jones
potential of Fig. \ref{fig:pwavek}. (a) Scaled partial cross-section
$\lambda\sigma_{11}$ as a function of $\lambda$. (b) $K$ matrix
element $\mathcal{K}_{11}$ as a function of $\lambda$. The blue
solid curves are the analytical results, and the yellow dashed curves
are determined by the $s$-wave only approximation Eq. (\ref{eq:Ksmat}).
The red crosses are results from a numerical calculation using the
same Lennard-Jones potential with the presence of SOC.\label{fig:pwavelambda}}

\end{figure}

Near $p$-wave resonances, the $p$-wave phase shift can no longer
be neglected. Neverthuless, a simplified fromula can be obtained in
the low-energy limit $k\rightarrow0$, where $\tan[\delta_{s}(k_{s})]\rightarrow-A_{s}\equiv-\tan[\delta_{s}(\sqrt{2}\lambda)]$
and $\tan[\delta_{p}(k_{p})]\rightarrow-A_{p}k^{2}$, and the $K$
matrix is given by,
\begin{equation}
\lim_{k\rightarrow0}\underline{\mathcal{K}}^{(+)}=\frac{1}{d}\left[\begin{array}{cc}
b_{11} & b_{12}k/\lambda\\
b_{21}k/\lambda & b_{22}k^{2}/\lambda^{2}
\end{array}\right].
\end{equation}
Here, $d$ and $b_{\tau'\tau}$ are all constants, which is given
by $d=1-2(\log\sqrt{2})A_{s}/\pi-2\lambda^{2}A_{p}/\pi+4\lambda^{2}A_{p}A_{s}[(\log\sqrt{2})-1]/\pi^{2}$,
$b_{11}=-A_{s}(1/2-\lambda^{2}A_{p}/\pi)$, $b_{12}=b_{21}=A_{s}(1/2+\lambda^{2}A_{p}/\pi)$
and $b_{22}=-A_{s}/4-\lambda^{2}A_{p}-\lambda^{2}A_{p}A_{s}(3/2-\log2)/\pi$.
When $|\lambda^{2}A_{p}|\gtrsim1$, $p$-wave scattering gives a significant
contribution as shown in Fig. \ref{fig:pwavelambda}, where Eq. (\ref{eq:Ksmat})
is no longer valid. Neverthuless, the threshold laws for cross-section
and $K$ matrix elements are valid for all situations as shown in
Figs. \ref{fig:swavek} and \ref{fig:pwavek}. Interestingly, the
elastic scattering rate $\propto k_{1}\sigma_{11}$ that determines
thermalization remains constant in the zero-energy limit, in contrast
to the vanishing $1/(\log k)^{2}$ rate without the presence of SOC.
We also remark here that, using the free-space pseudopotential including
the $p$-wave contribution ${\rm diag}[V_{1}^{{\rm fs}}(k,\rho),V_{0}^{{\rm fs}}(k,\rho),V_{1}^{{\rm fs}}(k,\rho)]$
will wrongly give a vanishingly small $K$ matrix due to a $\log(s)|_{s\rightarrow0}$
term in the denumerator of all the matrix elements of $\underline{\mathcal{K}}^{(+)}$,
reflecting the importance of the non-diagonal terms in the pseudopotnetial.

In summary, we have derived a pseudopotential in the COM frame with
the presence of SOC in 2D using a frame-transformation approach. Different
than the free-space pseudopotential, the s -wave scattering phase-shift
changes due to a SOC-induced energy shift. The frame-transformation
also introduces non-diagonal terms, which are also essential to reproduce
two-body scattering observables. We applied this pseudopotential with
the Lippmann-Schwinger equation to obtain the analytical scattering
matrix and compare it with a numerical scattering calculation with
finite-range potential. Our pseudopotential is valid even near $s$-
or $p$-wave resonances as long as $\lambda r_{{\rm vdW}}\ll1$, which
is usually well satisfied in ultracold quantum gases. Our results
indicate that, if we consider, if we consider $s$-wave only (which
usually implies near $s$-wave resonances), and the energy-dependency
of $a_{s}(k)$ is very weak (which usually implies a very broad resonance)
so that $a_{s}(\sqrt{2}\lambda)\approx a_{s}(0)$, the free-space
pseudopotential can give a good approximation, which gives the valid
regime of previous studies in Refs. \citep{ZhangweiPRA2012,ZhanglongPRA2013}.
On the other hand, if the energy-dependency of $a_{s}(k)$ is strong
or $p$-wave interaction is nonnegligible, our pseudopotential has
to be adopted to reproduce two-body scattering. Our approach can also
be easily applied in 3D and reproduce Eq. (11) of Ref. \citep{GuanQPRA2017},
which we will pursuit elsewhere. Our results are also useful for investigating
universal relations and Tan's contacts for SOC quantum gases in 2D
\citep{Peng2019} and might eventually be applied in many-body physics
studies.

\section*{Supplemental Material}

\subsection{Full analytical expression of $K$ matrix}

Here, we give the full analytical expression of $\underline{\mathcal{K}}^{(+)}$,

\begin{equation}
\underline{\mathcal{K}}^{(+)}=\left[\begin{array}{cc}
\mathcal{K}_{11} & \mathcal{K}_{12}\\
\mathcal{K}_{21} & \mathcal{K}_{22}
\end{array}\right]=\frac{1}{D}\left[\begin{array}{cc}
B_{11} & B_{12}\\
B_{21} & B_{22}
\end{array}\right],
\end{equation}
where \begin{widetext}

\begin{equation}
\begin{gathered}D=1+\frac{2}{\pi}t_{s}\left(\log\frac{k}{k_{s}}+\frac{\lambda}{k_{b}}\tanh^{-1}\frac{\lambda}{k_{b}}\right)+\frac{2}{\pi}t_{p}\left(\frac{\lambda^{2}}{k^{2}}-\frac{\lambda}{k_{b}}\tanh^{-1}\frac{\lambda}{k_{b}}\right)+\frac{4}{\pi^{2}}t_{s}t_{p}\left[\log k_{2}\log k_{3}-(\log k)^{2}\right]\\
+\frac{4}{\pi^{2}}t_{s}t_{p}\left[\frac{\lambda^{2}}{k^{2}}\left(\log\frac{k}{k_{s}}+\frac{\lambda}{k_{b}}\tanh^{-1}\frac{\lambda}{k_{b}}-1\right)+\left(2-\log\frac{k}{k_{s}}\right)\left(\frac{\lambda}{k_{b}}\tanh^{-1}\frac{\lambda}{k_{b}}\right)\right],
\end{gathered}
\end{equation}
\begin{equation}
B_{11}=t_{s}\frac{k_{1}}{2k_{b}}+t_{p}\frac{k_{2}}{2k_{b}}+\frac{t_{s}t_{p}}{\pi}\left(\frac{\lambda}{k_{b}}+\frac{\lambda}{k_{2}}-\frac{\lambda}{k_{b}}\log\frac{k}{k_{s}}-\log\frac{kk_{s}}{k_{2}^{2}}\right),
\end{equation}
\begin{equation}
B_{12}=-t_{s}\frac{k}{2k_{b}}+t_{p}\frac{k}{2k_{b}}+\frac{2}{\pi}t_{s}t_{p}\frac{k}{2k_{b}}\left(\log\frac{k}{k_{s}}+\frac{\lambda^{2}}{k^{2}}\right),
\end{equation}

\begin{equation}
B_{21}=-t_{s}\frac{k}{2k_{b}}+t_{p}\frac{k}{2k_{b}}+\frac{2}{\pi}t_{s}t_{p}\frac{k}{2k_{b}}\left(\log\frac{k}{k_{s}}+\frac{\lambda^{2}}{k^{2}}\right),
\end{equation}

\begin{equation}
B_{22}=t_{s}\frac{k_{2}}{2k_{b}}+t_{p}\frac{k_{1}}{2k_{b}}-\frac{t_{s}t_{p}}{\pi}\left(\frac{\lambda}{k_{b}}+\frac{\lambda}{k_{1}}-\frac{\lambda}{k_{b}}\log\frac{k}{k_{s}}+\log\frac{kk_{s}}{k_{1}^{2}}\right),
\end{equation}

\end{widetext} with the notations $t_{s}\equiv\tan\left[\delta_{s}(k_{s})\right]$,
$t_{p}\equiv\tan[\delta_{p}(k_{p})]$, $k_{s}=\sqrt{k^{2}+2\lambda^{2}}$,
$k_{p}=k$, $k_{b}=\sqrt{k^{2}+\lambda^{2}}$, $k_{1}=k_{b}+\lambda$
and $k_{2}=k_{b}-\lambda$.

\subsection{Numerical method}

We carry out a numerical calculation with finite range Lennard-Jones
potentials $U(\rho)$ to verify our analytical results, using a similar
procedure as Ref. \citep{WangjiaPRA2018}. We expand the rescaled
Hamiltonian $h\equiv\rho^{1/2}H_{{\rm rel}}\rho^{-1/2}$ with the
channel functions that leads to a set of coupled differential equations
for $\underline{u}=\rho^{1/2}\underline{R}^{{\rm SO}}$: $\underline{h}\underline{u}=E\underline{u}$,
where $\underline{h}=\underline{h}^{{\rm fs}}+\underline{V}^{{\rm SO}}$
has matrix elements
\begin{equation}
h_{\nu^{\prime}\nu}^{{\rm fs}}=\left[-\frac{\hbar^{2}}{\mu_{2b}}\left(\partial_{\rho}^{2}-\frac{m_{\ell}^{2}-\frac{1}{4}}{\rho^{2}}\right)+U(\rho)\right]\delta_{\nu\text{\textquoteright}\nu}
\end{equation}
and
\begin{equation}
V_{\nu^{\prime}\nu}^{{\rm SO}}=\frac{\hbar^{2}k_{\mathrm{soc}}}{2\mu_{2b}}\left[\Sigma_{\chi^{\prime}\chi}^{(+)}p_{m_{\ell}^{\prime}m_{\ell}}^{(-)}+\Sigma_{\chi^{\prime}\chi}^{(-)}p_{m_{\ell}^{\prime}m_{\ell}}^{(+)}\right],
\end{equation}
where $p_{m_{\ell}^{\prime}m_{\ell}}^{(\pm)}=\mp(\partial_{\rho}-1/2\rho\pm m_{\ell}/\rho)\delta_{m_{\ell}^{\prime},m_{\ell}\pm1}$.
Here we define $\Sigma_{\chi^{\prime}\chi}^{(\pm)}=\left\langle \chi^{\prime}\left|s_{1}^{\pm}-s_{2}^{\pm}\right|\chi\right\rangle /2$
with $s_{n}^{+}$ ( $s_{n}^{-}$ ) beign the raising (lowering) operator
for the spin state of the $n$'th particle, which can be obtained
explicitly as 
\begin{equation}
\begin{array}{cc}
\Sigma_{\chi^{\prime}\chi}^{(\pm)}= & \sum_{m_{1}m_{2}}C_{s_{1}m_{1};s_{2}m_{2}}^{Sm_{S}}(a_{1\pm}C_{s_{1}m_{1}\pm1;s_{2}m_{2}}^{S^{\prime}m_{S}^{\prime}}\\
 & -a_{2\pm}C_{s_{1}m_{1};s_{2}m_{2}\pm1}^{S^{\prime}m_{S}^{\prime}})/2
\end{array}
\end{equation}
with $a_{n\pm}=\sqrt{s_{n}(s_{n}+1)-m_{n}(m_{n}\pm1)}$ and $C_{s_{1}m_{1};s_{2}m_{2}}^{Sm_{S}}$
being Clebsch-Gordan coefficients. For numerical calculations, one
can solve the multichannel radial Schr{\"o}dinger equations by propagating
wavefunction matrix $\underline{u}$ or equivalently the logarithmic
derivative matrix $\underline{\mathcal{L}}=\underline{u}'\underline{u}^{-1}$
to a large enough distances $\rho_{{\rm max}}\gg\rho_{0}$. The $K$
matrix can be obtained via $\underline{\mathcal{K}}=(\underline{\mathcal{L}}\underline{g}-\underline{g}')^{-1}(\underline{\mathcal{L}}\underline{f}-\underline{f}')$,
where $\underline{f}=\sqrt{\rho}\underline{F}$ and $\underline{g}=\sqrt{\rho}\underline{G}$.
The numerical results are shown as symbols in all figures in the main
text, showing perfect agreement in the range of $k$ and $\lambda$
considered.

\bibliographystyle{apsrev4-1}
\bibliography{Refs2DSOC}

\end{document}